\documentclass[12pt]{article}
\usepackage{amsmath}
\usepackage[cp1251]{inputenc}
\usepackage{bm}
\usepackage{color}
\usepackage{caption}
\usepackage{subcaption}
\usepackage{graphicx}
\usepackage{braket}

\newcommand{\mi}
{\mathrm{i}}

\begin{document}
\title{Evaluation of variational quantum states entanglement on a quantum computer by the mean value of spin}

\author{
Kh. P. Gnatenko$^1$\footnote{khrystyna.gnatenko@gmail.com}, \\
Ivan Franko National University of Lviv,\\
Professor Ivan Vakarchuk Department for Theoretical Physics,\\
12, Drahomanov St., Lviv, 79005, Ukraine \and
 SoftServe Inc., 2d Sadova St., 79021 Lviv, Ukraine}

\maketitle

\begin{abstract}
The geometric measure of entanglement of variational quantum states is studied on the basis of its relation with the mean value of spin.  We examine n-qubit quantum states prepared by a variational circuit with a layer formed by the rotational gates and two-qubit controlled phase gates.  The variational circuit is a generalization of that used for preparing quantum Generative Adversarial Network states.   The entanglement of a qubit with other qubits in the variational quantum states is determined by the angles of rotational gates that act on the qubit and qubits entangled with it by controlled phase gates and also their parameters. In the case of one layer variational circuit, the states can be associated with graphs with vertices representing qubits and edges corresponding to two-qubit gates. The geometric measure of entanglement of a qubit with other qubits in the quantum graph state depends on the properties of the vertex that represents it in the graph, namely it depends on the vertex degree.  The dependence of the geometric measure of entanglement of variational quantum states on their parameters is quantified on IBM's quantum computer.

\end{abstract}

\section{Introduction}

Variational quantum circuits have received much attention because of their wide usage in various quantum algorithms among them Quantum Machine Learning, error correction, algorithms for solving optimization problems, and many others (see, for instance, \cite{Zoufal,Cerezo,Yuxuan,Bravo,Xiaosi,Moll,Wecker} and references therein). A critical and festinating resource in quantum computing is entanglement \cite{Horodecki,Shimony,Pierro,Behera}. The geometric measure of entanglement of quantum states has a clear definition that is based on the geometric idea of computing the minimal distance between the entangled state and a set of non-entangled states.
The measure of entanglement can be found as  $E(\mid\psi\rangle)=\min_{\mid\psi_s\rangle}d^2_{FS}(|\psi_s\rangle, |\psi\rangle)$,
where $\mid\psi\rangle$  is an entangled state,   $\mid\psi_s\rangle$ is a   set of non-entangled states, $d_{FS}=\sqrt{1-|\langle\psi|\psi_s\rangle|^2}$ is the
 Fubini-Study distance between the states $\mid\psi_s\rangle$, $\mid\psi\rangle$  \cite{Shimony}. The procedure of finding the minimal distance requires a lot of computational resources. In the paper, \cite{Frydryszak} it was found that to detect the geometric measure of entanglement of a spin with an arbitrary quantum system in a pure state it is sufficient to calculate the mean value of the spin. Namely, the geometric measure of entanglement of a  spin with quantum system that are in a state $\mid\psi\rangle=a\mid\uparrow\rangle\mid\Phi_1\rangle+b\mid\downarrow\rangle\mid\Phi_2\rangle,$
can be found as
\begin{eqnarray}
E(\mid\psi\rangle)=\frac{1}{2}(1-\sqrt{\langle{\bm \sigma}\rangle^2}).\label{ent}
\end{eqnarray}
Here $\mid\Phi_1\rangle$, $\mid\Phi_2\rangle$ are states of a quantum system ($\langle\Phi_i\mid\Phi_i\rangle=1$) that in general case can be nonorthogonal, $a$, $b$ are constants,
\begin{eqnarray}
\sqrt{\langle{\bm \sigma}\rangle^2}=\sqrt{\langle{ \sigma^x}\rangle^2+\langle{ \sigma^y}\rangle^2+\langle{ \sigma^z}\rangle^2},
 \end{eqnarray}
 $\sigma^x$, $\sigma^y$, $\sigma^z$ are Pauli matrixes \cite{Frydryszak}. It is important to mention that relation (\ref{ent}) opens a possibility to quantify  the geometric measure of entanglement of quantum states on a quantum devices.

We study the geometric measure of entanglement of n-qubit quantum states prepared by a variational circuit with a layer formed by the rotational gates and controlled phase gates on the basis of analytical calculations and programming on IBM's quantum computer and IBM's quantum simulator.

\section{Detection of the geometric measure of entanglement of variational quantum states on IBM's quantum computer}

\begin{figure}[!!h]
 \begin{center}
\includegraphics[scale=0.55]{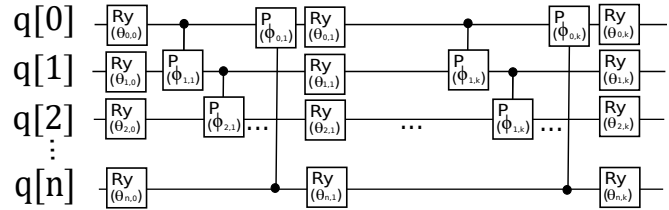}
\caption{Variational quantum circuit of a depth k.  The layer formed by the rotational gates $RY(\theta_{i,j})$ and controlled phase gates $CP(\varphi_{i,j})$. }\label{fig1}
 \end{center}
\end{figure}

 Let us consider variational quantum circuit with a layer formed by the rotational gates $RY(\theta_{i,j})$ and controlled phase gates $CP(\varphi_{i,j})$  presented in Fig. \ref{fig1}.
Note that in the particular case of $\varphi_{i,j}=\pi$  the  variational n+1-qubit form corresponds to that used as a quantum generator in the quantum Generative Adversarial Networks (qGANs) \cite{Zoufal}.

\begin{center}
\begin{figure}[!!h]
 \begin{center}
\includegraphics[scale=0.55]{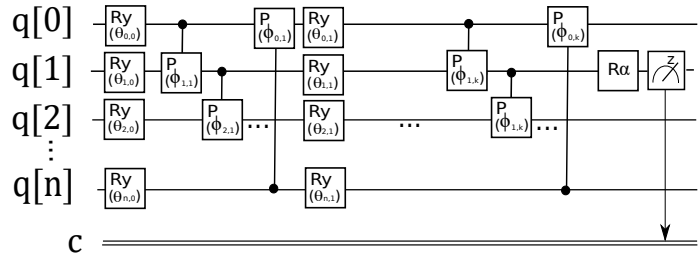}
\caption{Quantum protocol for detection of the geometric measure of entanglement  of qubit $q[1]$ with other qubits in a state  prepared by the variational quantum circuit with the depth $k$.}\label{fig2}
 \end{center}
\end{figure}
\end{center}

To detect the geometric measure of entanglement of a qubit with other qubits in a state prepared with a variational circuit we quantify the mean value of operators $\sigma^x$, $\sigma^y$, $\sigma^z$ in the corresponding variational quantum state. For instance, to evaluate the entanglement of qubit $q[1]$ with other qubits in the state we consider quantum protocol presented in Fig. \ref{fig2}
In the quantum protocol it is taken into account that
$\langle \psi \vert \sigma^x\vert\psi\rangle=\langle \tilde\psi^y \vert \sigma^z\vert \tilde\psi^y\rangle=\vert \langle \tilde\psi^y \vert 0 \rangle \vert^2-\vert\langle \tilde\psi^y \vert 1 \rangle \vert^2$, $\langle \psi \vert \sigma^y\vert\psi\rangle=\langle \tilde\psi^x \vert \sigma^z\vert \tilde\psi^x\rangle=\vert \langle \tilde\psi^x \vert 0 \rangle \vert^2-\vert\langle \tilde\psi^x \vert 1 \rangle \vert^2$,  where $\vert\tilde\psi^x\rangle=\exp({-\mi{\pi}\sigma^x/{4}})\vert\psi\rangle$, $\vert\tilde\psi^y\rangle=\exp({\mi{\pi}\sigma^y/{4}})\vert\psi\rangle$. Therefore to quantify the mean value $\langle\sigma^x \rangle$ the $R_{\alpha}=RY(-\pi/2)$ gate has to be applied  to the state of a qubit before the measurement in the standard basis.
To detect the mean value $\langle\sigma^y \rangle$ the $R_{\alpha}=RX(\pi/2)$ gate has to be used. Mean value of $\sigma^z$ operator can be calculated, using  the results of measurement in the standard basis $\langle \psi \vert \sigma^z\vert\psi\rangle=\vert \langle \psi \vert 0 \rangle \vert^2-\vert\langle \psi \vert 1 \rangle \vert^2$.

 It is worth mentioning that quantum states generated by variational quantum circuit with the depth $k=1$ are graph states. These states can be associated with undirected graphs $G(E,V)$. The qubits  are represented by vertices $V$ in the graph  and the edges $E$ correspond to two-qubit gates. The geometric measure of entanglement of a qubit with other qubits in the graph  state
$\mid G\rangle=\prod_{(a,b)\in E} CP_{ab}(\phi)\prod_i RY_i(\theta)\mid 0\rangle^{\otimes n},$ (here $RY_i(\theta)$, $CP_{ij}(\phi)$ are rotational and controlled phase gates acting on the states of qubits $q[i]$, $q[j]$)
 depends on the graph properties \cite{Gnatenko1,Gnatenko2}.
Namely, the entanglement of qubit $q[l]$ with other qubits  in n-qubit graph state  depends on the the degree of vertex $n_l$  that represents it.  Performing  analytical calculations, on the basis of the result (\ref{ent}), one finds \cite{Gnatenko1}
\begin{eqnarray}
		E_l =  \frac{1}{2}-\frac{1}{2}\sqrt{\sin^2\theta \left(\cos^2 \frac{\phi}{2}+\sin^2 \frac{\phi}{2}\cos^2 \theta \right)^{n_l} + \cos^2 \theta }.\label{1}
	\end{eqnarray}

 Using relation of the geometric measure of the entanglement with the mean value of spin (\ref{ent}), we obtain  the entanglement of qubit $q[1]$ with other qubits in the quantum Generative Adversarial Network state
 prepared by the $n+1$ qubit variational quantum circuit  Fig. \ref{fig1} with depth $k=1$, and  $\phi_{i,1}=\pi$. It
reads
\begin{eqnarray}
E(\theta_{0,0},\theta_{1,0},\theta_{2,0})=
\frac{1}{2}\left(1-\sqrt{\cos^2\theta_{0,0}\cos^2\theta_{2,0}\sin^2\theta_{1,0}+\cos^2\theta_{1,0}}\right).
\end{eqnarray}
Note that the entanglement of qubit $q[1]$ with other qubits in the state depends only on the parameters $\theta_{0,0}$, $\theta_{1,0}$, $\theta_{2,0}$ of $RY$ gates  that act on the qubits entangled with $q[1]$ by $CZ_{01}$, $CZ_{12}$ gates and do not depend on other parameters of the variational circuit.

\begin{figure}[!!h]
\includegraphics[scale=0.2, angle=0.0, clip]{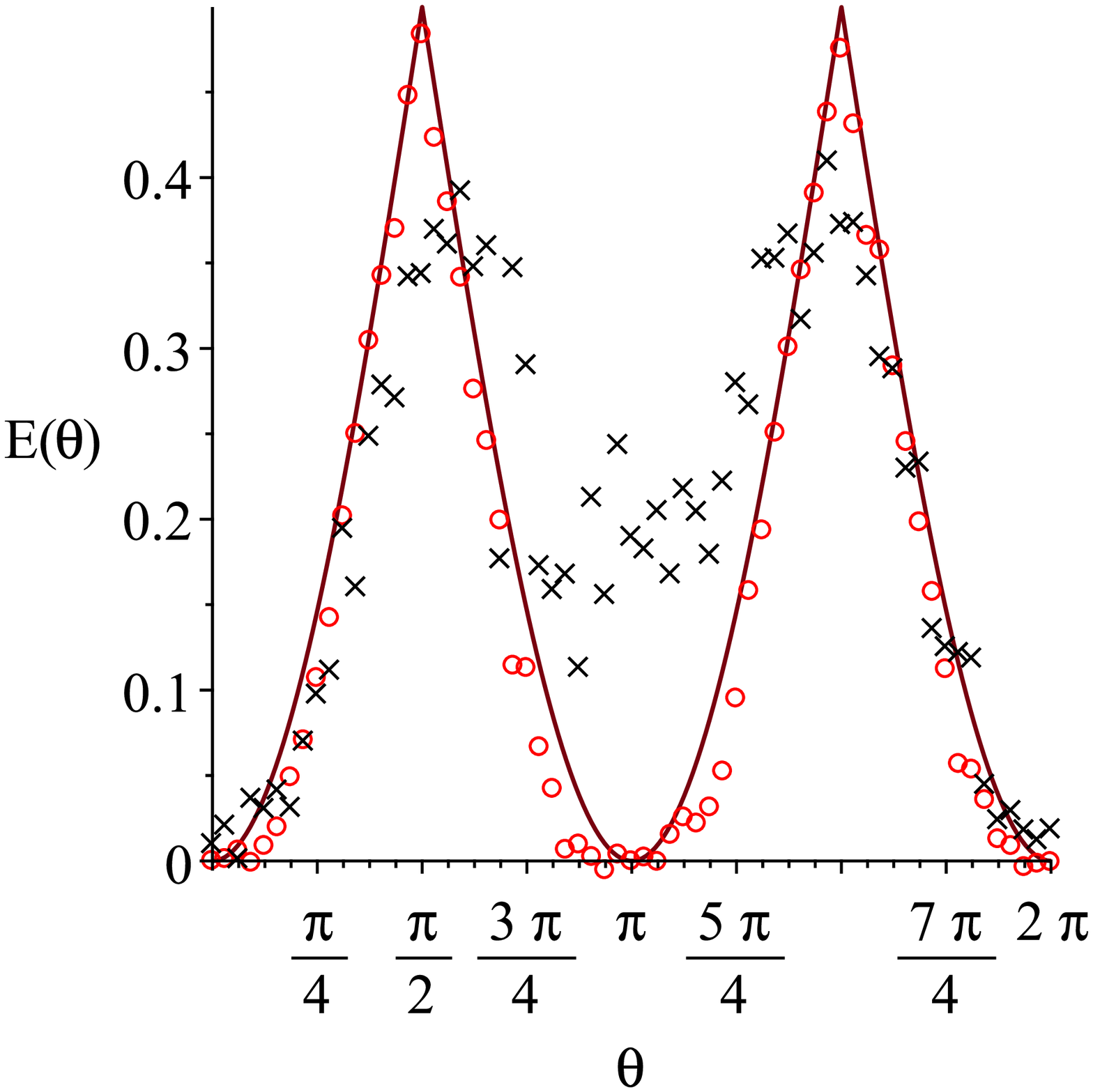}
\includegraphics[scale=0.2, angle=0.0, clip]{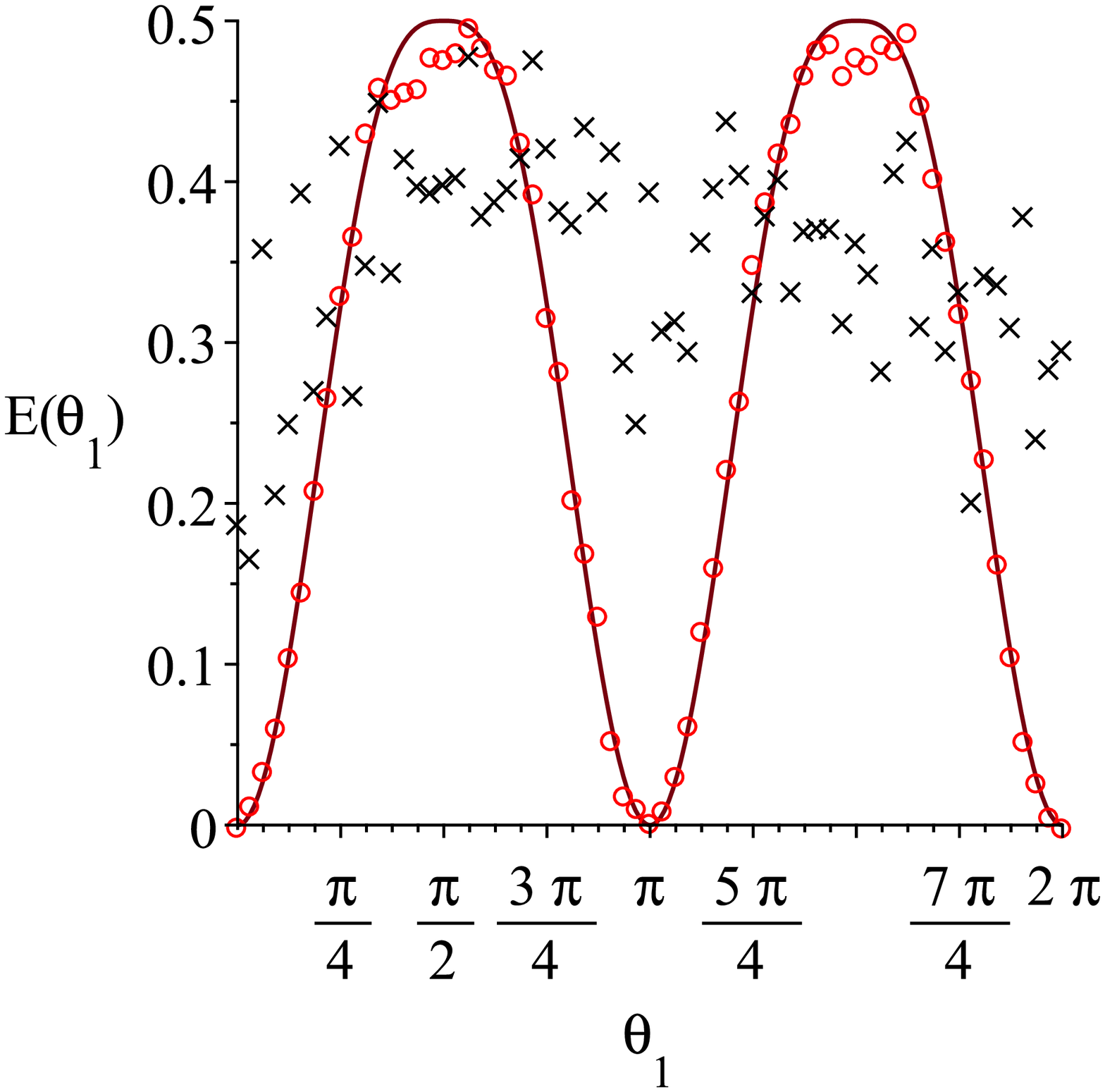}
\includegraphics[scale=0.2, angle=0.0, clip]{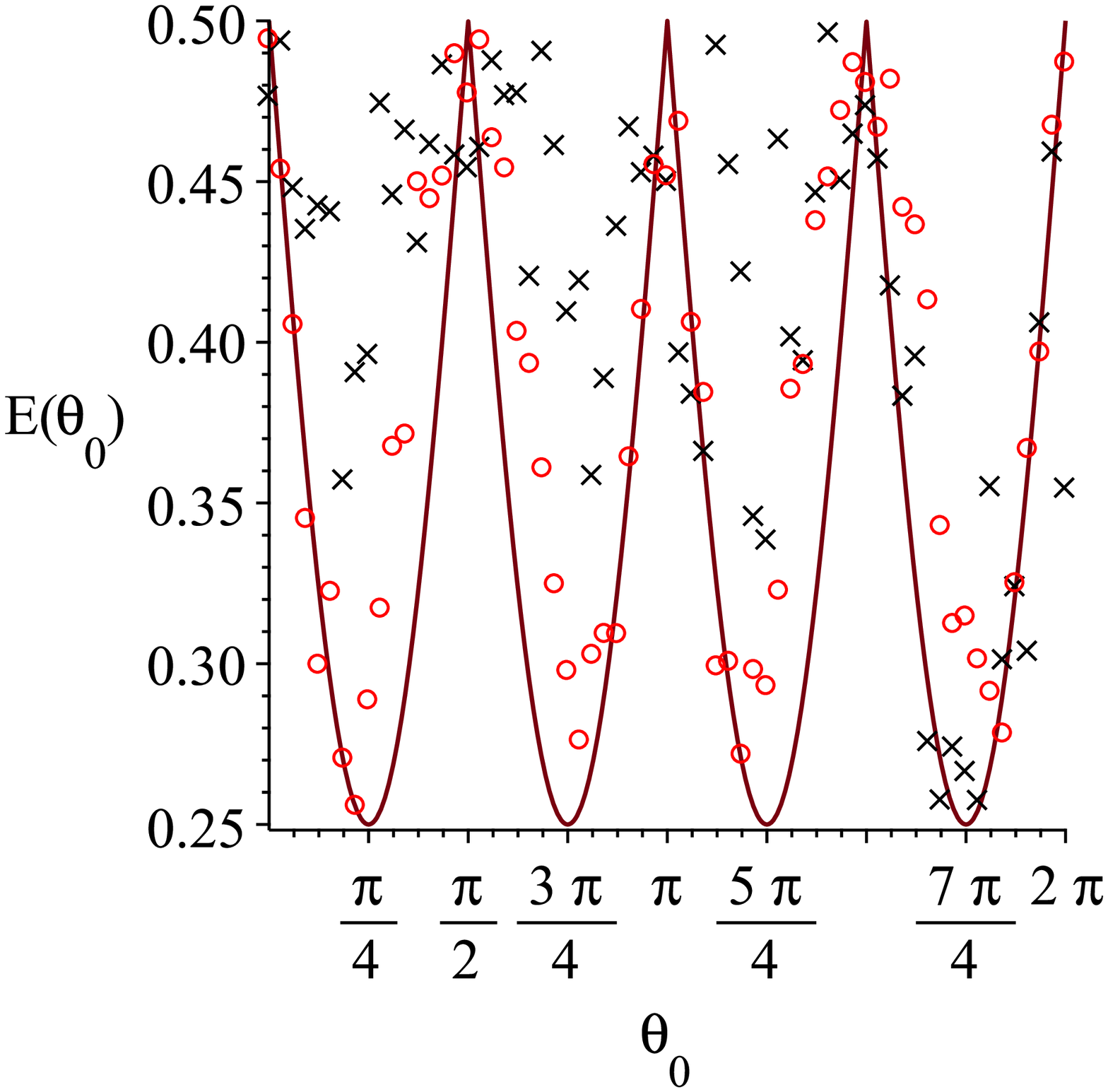}
\caption{Results of calculations of entanglement of qubit $q[1]$ with other qubits in variational quantum states  (a) for different values of $\theta_{i,0}=\theta$, $\phi_{i,1}=\pi$ and $k=1$   (left plot);  (b) for different values of $\theta_{i,1}=\theta_{1}$ and $\theta_{i,0}=\pi/2$, $\phi_{i,1}=\phi_{i,2}=\pi$, $k=2$  (middle plot); (c)  for different values of $\theta_{i,0}=\theta_{0}$ and $\theta_{i,1}=\pi/2$, $\phi_{i,1}=\phi_{i,2}=\pi$, $k=2$ (right plot), obtained  on  ibmq-manila (marked by black crosses), ibmq-qasm-simulator (marked by red circles), and analytical results (line).} \label{fig5}
\end{figure}

To study the geometric measure of entanglement of the quantum Generative Adversarial Network states on a quantum device we realized quantum protocol Fig. \ref{fig2}  on  ibmq-manila and ibmq-qasm-simulator for $k=1$, $\phi_{i,1}=\pi$,  $\theta_{i,0}=\theta$  changing from $0$ to $2\pi$ with the step $\pi/32$. The results of calculations  are presented in Fig. \ref{fig5} (a).

 Let us also examine the geometric measure of entanglement  of qubit $q[1]$ with other qubits in the quantum Generative Adversarial Network state generated by variational circuit with depth  $k=2$. In the case of  $\theta_{i,0}=\pi/2$ the entanglement reads
\begin{eqnarray}
E(\theta_{0,1},\theta_{1,1},\theta_{2,1})=\frac{1}{2}(1-\frac{1}{4}\vert\cos(\theta_{0,1}+\theta_{1,1}-\theta_{2,1})+\nonumber\\+\cos(\theta_{0,1}-\theta_{1,1}+\theta_{2,1})
+\cos(-\theta_{0,1}+\theta_{1,1}+\theta_{2,1})+\cos(\theta_{0,1}+\theta_{1,1}+\theta_{2,1})\vert).
\end{eqnarray}
We realized protocol Fig. \ref{fig2} for  $k=2$,  $\theta_{i,0}=\pi/2$, $\theta_{i,1}=\theta_1$, $\phi_{i,1}=\phi_{i,2}=\pi$  and  $\theta_1$ changing from $0$ to $2\pi$ with the step $\pi/32$  on quantum device   ibmq-manila and ibmq-qasm-simulator.
The results of quantum calculations  and the theoretical result for the entanglement in this case  $E(\theta_{1})=(1-\vert\cos^3\theta_{1}\vert)/2$ are plotted in Fig. \ref{fig5} (b). Quantum protocol Fig. \ref{fig2} was also implemented for, $k=2$,  $\theta_{i,0}=\theta_0$, $\theta_{i,1}=\pi/2$, $\phi_{i,1}=\phi_{i,2}=\pi$  and  $\theta_0$ changing from $0$ to $2\pi$ with the step $\pi/32$ on the quantum computer and simulator. The results of quantum calculations and the analytical result $E(\theta_{0})=(2-\vert\sin(2\theta_{0})\vert)/2$ are presented in Fig. \ref{fig5} (c).

In addition, to detect the dependence of the geometric measure of entanglement on the parameters of phase gates we realized quantum protocol Fig. \ref{fig2}  for different values of $\phi_{i,1}=\phi$ changing from $0$ to $4\pi$ with the step $\pi/32$ and  $\theta_{i,0}=\pi/2$ (the depth  is $k=1$)   Fig. \ref{fig2} (a) and for different values of $\phi_{i,1}=\phi_1$, changing from $0$ to $4\pi$ with the step $\pi/32$,  $\theta_{i,0}=\pi/2$, $\phi_{i,2}=\pi$ (the  depth  is $k=2$)  Fig. \ref{fig2} (b). The results of quantum calculations and the analytical results $E(\phi)=(1-\cos^2(\phi/2))/2$,  $E(\phi_1)=(1-\vert\cos(\phi_1)(1+\cos(\phi_1))\vert)/2$   are presented in Fig. \ref{fig2}, cases (a) and (b), respectively.  Note that for $k=1$, the results of quantum calculations are in good correspondence  with the theoretical ones. Not so good agreement of the results was obtained for circuits with the depth $k=2$ because of accumulating of the gate errors.

\begin{figure}[!!h]
\includegraphics[scale=0.3, angle=0.0, clip]{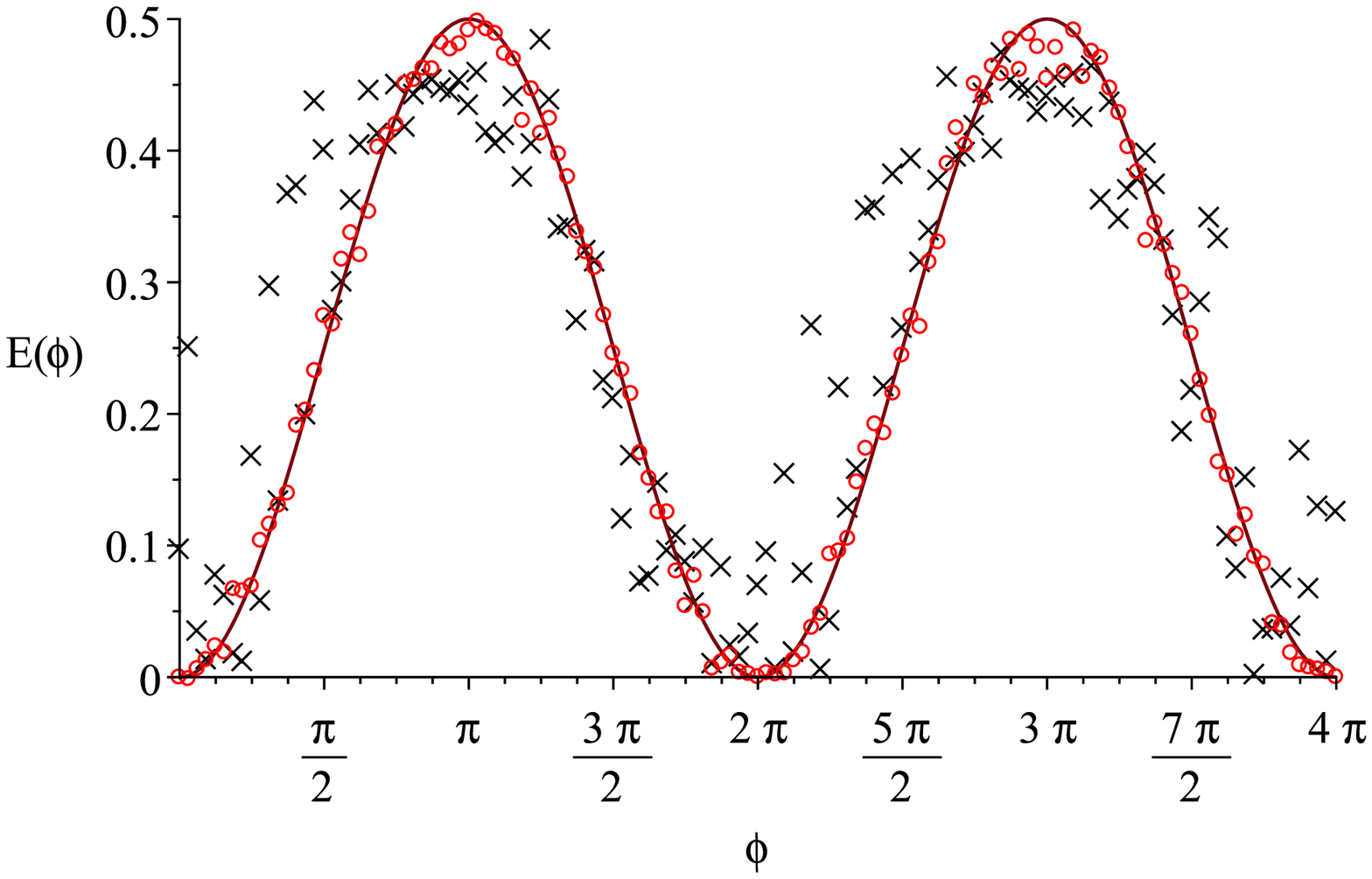}
\includegraphics[scale=0.32, angle=0.0, clip]{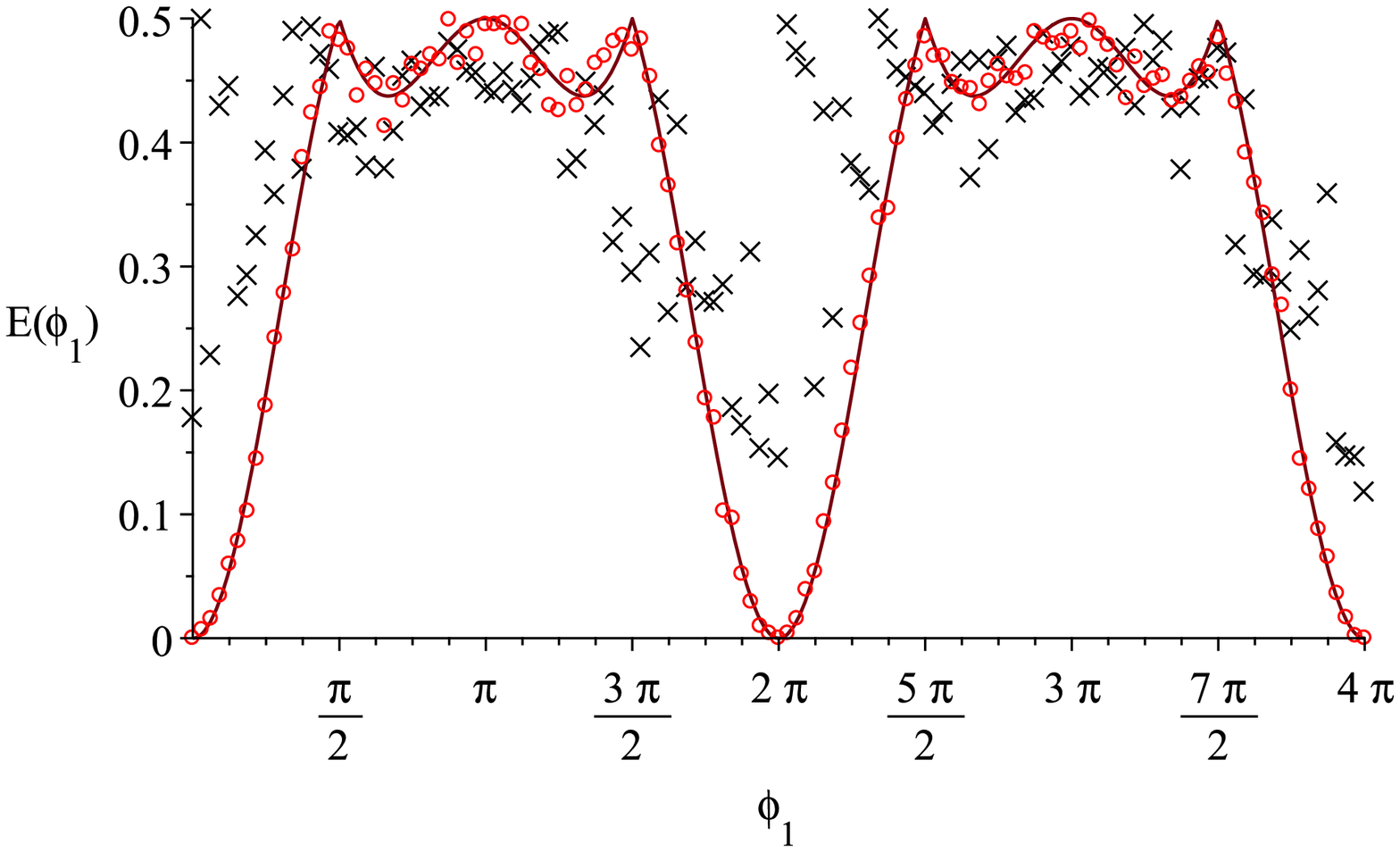}
\caption{Results of calculations of entanglement of qubit $q[1]$ with other qubits in  variational quantum states   (a) for different values of $\phi_{i,1}=\phi$, $\theta_{i,0}=\pi/2$, and the depth  $k=1$   (left plot);  (b) for different values of $\phi_{i,1}=\phi_1$  and $\theta_{i,0}=\pi/2$, $\phi_{i,2}=\pi$, and the depth  $k=2$  (right plot), obtained on the on  ibmq-manila (marked by black crosses),  ibmq-qasm-simulator (marked by red circles), and analytical results (line) } \label{fig6}
\end{figure}

\section{Conclusions}
We have studied the geometric measure of entanglement of a variational quantum states prepared by   rotational gates and entangled blocks formed by controlled phase gates Fig. \ref{fig1}. The studies have been done on the basis of the relation of the entanglement with the mean value of spin (\ref{ent}). In the case of variational circuit of one layer the variational quantum state can be associated with graph and the the geometric measure of entanglement of a qubit with other qubits in the variational quantum state with parameters $\theta_{i,0}=\theta$, $\phi_{i,0}=\phi$ is related with the degree of vertex representing it in the graph (\ref{1}).

The entanglement of a qubit with other qubits in the variational quantum state is determined by the angles of rotational gates in the variational circuit that act on the qubits entangled with it by controlled phase gates and their parameters. The relation of the geometric measure of the entanglement with mean value of spin opens a possibility to quantify the entanglement on quantum devices, realizing quantum protocol Fig. \ref{fig2}. On the basis of the relation the dependencies of the variational quantum states entanglement on the  parameters of the variational quantum circuit have been calculated on  IBM's quantum computer  ibmq-manila and ibmq-qasm-simulator Figs. \ref{fig5}, \ref{fig6}.

\section*{Acknowledgements}  The author thanks Prof. Tkachuk V. M. for valuable discussions in the filed of studies.

\end{document}